# BEAM TRANSPORT IN TOROIDAL MAGNETIC FIELD*

N. Joshi, M. Droba, O. Meusel, U. Ratzinger, IAP, Frankfurt am Main, Germany

## Abstract

The concept of a storage ring with toroidal magnetic field was presented in the two previous EPAC conferences. Here we report the first results of experiments performed with beam transport in toroidal magnetic fields and details of the injection system. The beam transport experiments were carried out with 30 degree toroidal segments with an axial magnetic field of 0.6T. The multi turn injection system relies on a transverse injection coil together with an electric kicker system.

## Introduction

The magnetic field configuration of a figure-8 type stellarator is considered to be a promising concept for intense ion beam accumulation and storage in the 100$keV$ to a few $MeV$ energy range provides the confinement of ion beam [1] [2]. The 3d-greometry of the ring leads to closed magnetic surfaces. Fig. 1 shows the structure of magnetic surfaces along the whole ring and coil arrangement.

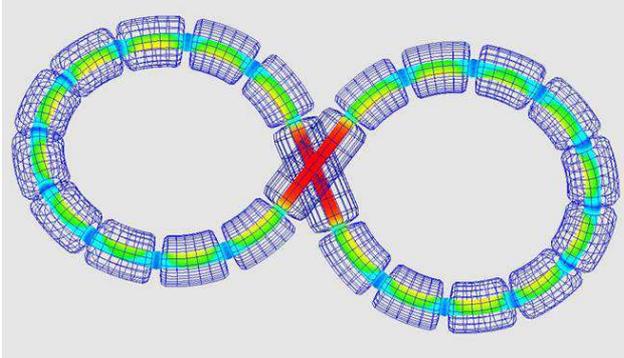

Figure 1: Schematic structure of ring built with identical toroidal segments. Colour coding on the magnetic surface indicates the magnetic field strength distribution.

The transverse **R**x**B** beam drift arising from curved magnetic fields can be compensated by the figure-8 type geometry. The Brillouin limit sets the maximum current limit, given by

$$n_B = \frac{\varepsilon_0 B^2}{2m}, \qquad (1)$$

which is ~$10^{16} m^{-3}$ for B=5T. For 150$keV$ beam energy current density is 5.6$A/cm^2$.

To investigate the beam transport into toroidal magnetic fields, experiments with two toroidal segments producing 0.6$T$ field were designed. An adequate beam energy range in this case is 4-20$keV$ protons.

For transport simulations in toroidal segments a PIC code was written. Fig. 2 shows the simulated magnetic array. At the input different beam distributions were tested.

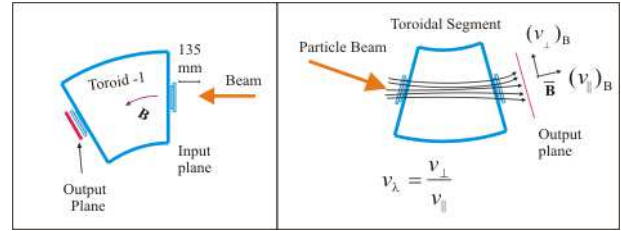

Figure 2: Schematic view used for simulations and experimental setup.

To study the phase-space distribution the velocity ratio $v_\lambda$ was introduced as,

$$v_\lambda = \left(\frac{v_\perp}{v_\parallel}\right)_B, \qquad (2)$$

where the transversal, and parallel velocities were defined with respect to the local magnetic field direction. This practice is commonly used in plasma physics. This gives a helpful condition on the beam transport: The particles are defined to form a beam when the ratio $v_\lambda < 0.1$. This gives the "good beam" condition, and all the particles with $v_\lambda > 0.1$ are called 'trapped particles'.

### Beam transport experiments in a 30° segment

Fig 3 shows the photograph of the experimental setup that was used to study the beam dynamics along a toroidal magnetic field.

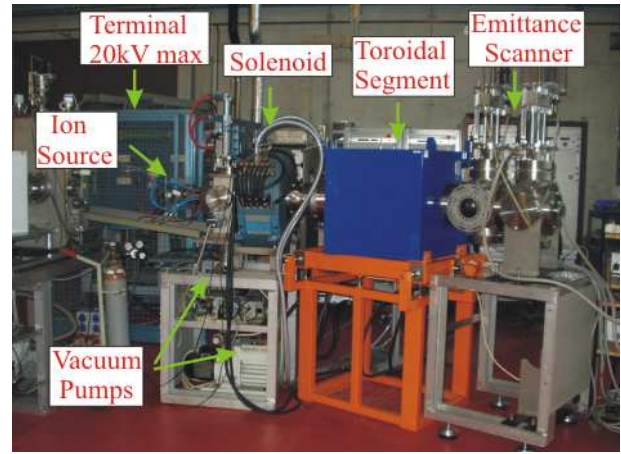

Figure 3: Photograph of the experimental setup for beam transport in a single toroidal segment.

A volume type ion source was used. The solenoid (0.72$T$ max on axis) allowed to match the phase-space distribution of the proton beam for injection into the toroidal segment. Downstream of the toroidal segment the emittance scanner (grid-slit device) was installed.

Table 1 lists important properties of a single toroidal segment. Two identical segments were manufactured. They are not magnetically shielded, and show extended fringing fields which have to be included in all simulations and measurements.

Table 1: Toroidal Segment Specifications

| Property | Value |
| --- | --- |
| Major radius | 1300 mm |
| Minor radius | 100 mm |
| Arc Angle of segment | 30 degree |
| Central arc length | 680 mm |
| Power Supply | 500A, 120V |
| Maximum magnetic on axis field (@480 A) | 0.6 T |

In a first experimental step the characterization of the ion source was carried out to find the optimum settings for a maximum proton fraction [3]. The maximum of 3.08$mA$ proton beam at 10$keV$ was successfully extracted and the phase-space distribution was measured by the emittance scanner. The measured distribution was used to simulate the transport into the solenoid. Simulated and measured data were in good agreement.

Toroidal beam transport at different magnetic fields (0.4-0.625$T$) was carried out using different injection energies (4-12$keV$) in the second experimental stage. Due to the toroidal field configuration with fringe fields and distances between slit and grid, limitations were imposed on measurements. It was found for certain settings that a good beam signal was registered, e.g. at 12$keV$, 0.4$T$, 0.5$T$ and 0.6$T$ fields. When one species of beam hits the walls or when large transversal velocities arise high noise levels come up in some cases.

Fig 4(A) shows the input phase-space distribution and fig. 4(B) the simulated phase-space distribution for 12keV proton beam. The beam consisted of 45% $H^+$, 54% $H_3^+$ and of a negligible $H_2^+$ fraction for these settings. Fig 4(C) shows the measured phase-space distribution downstream of the toroidal segment. When geometrical conditions of slit-grid along with "good beam" condition ($v_\lambda<0.1$) applied on the full phase-space simulated in fig. 4(B), we get the graph shown in fig 4(D). The velocity ratio is colour coded as 3$^{rd}$ dimension. The measured beam size (r~10$mm$ for $H^+$, r~30$mm$ for $H_3^+$) and output angle are in good agreement with simulations.

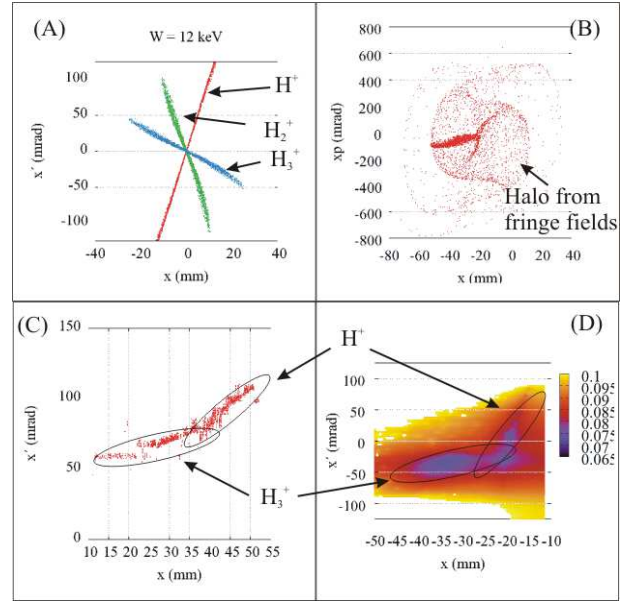

Figure 4: (A) Measured Input phase-space of distribution of beam, Red=$H^+$, Green=$H_2^+$, Blue=$H_3^+$, (B) Complete phase-space distribution at output of segment, (C) Measured phase-space distribution with emittance scanner, (D) Simulated phase space distribution chopped off with implementing emittance scanner acceptance with velocity ratio colour coded

*Injection System*

For injection experiments a special field configuration was designed. Fig. 5 shows the general layout of the scheme. The setup is an extension of experiments with single segment transport experiments described in the previous section. The second segment is to be mounted on the circular perimeter to represent a 60° ring section. Two beams will represent the circulating beam and the injected beam respectively.

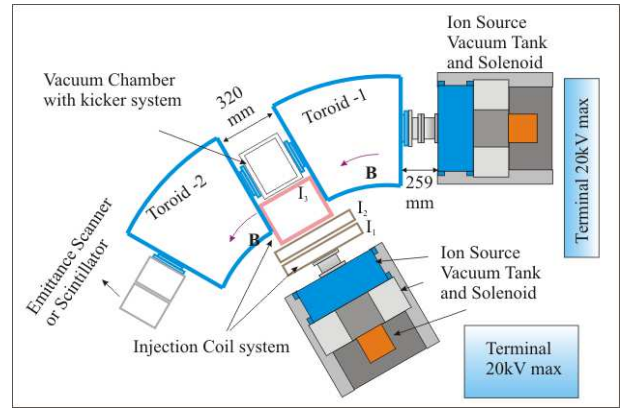

Figure 5: Layout of injection system.

As both segments are not shielded magnetically the auxiliary coil was designed for a smooth injection of the proton beam into the second toroidal segment. The coils

$I_1$, $I_2$ has large aperture (300mm) thus 2-dimensional flexible adjustment is possible. This provides the longitudinal momentum gain required to impose the input phase condition on the injected beam. Fig. 6 shows the transversal acceptance of the injector. The input distribution (at the entarance to $I_1$) is colour coded with respect to velocity ratio ($v_\lambda$) which represents the efficiency of the injection scheme.

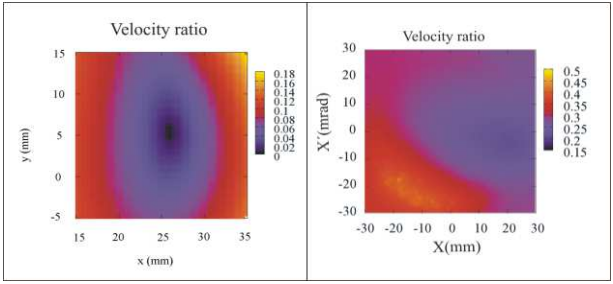

Figure 6: Acceptance at the input plane of the transversal beam injector.

It can be seen that a 10*mm* beam with 10*keV* energy and 2*mA* beam current can be successfully injected into the second segment. The self field effects and mechanical error studies were included in the simulations.

*Kicker system*

As shown in fig. 7 the auxiliary field configuration introduces new magnetic field lines which guide the beam into the second segment. But the particles injected on these new field lines tend to be lost from the ring as flux conservation does not allow auxiliary field lines to join the main ring flux.

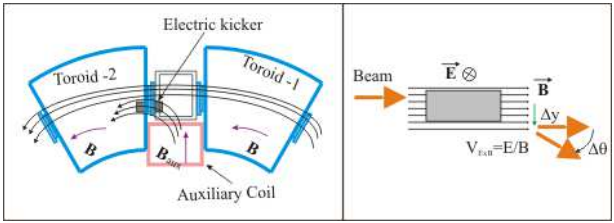

Figure 7: Position of electric kicker (left) and definition of ΔΘ (right).

Thus an electric kicker system (1.2*kV/cm*) which makes use of the **ExB** drift was proposed to transfer the injected beam on the ring field lines.

The applied drift velocity is given by,

$$V_E = \frac{E}{B}, \quad (4)$$

which is independent of mass. It can be derived that the minimum output angle ΔΘ (see fig. 7) of the beam depends on the length of the electric plates which is to be synchronized with respect to the energy.

$$L = \frac{(2n-1)\pi v_z m}{qB} \quad (5)$$

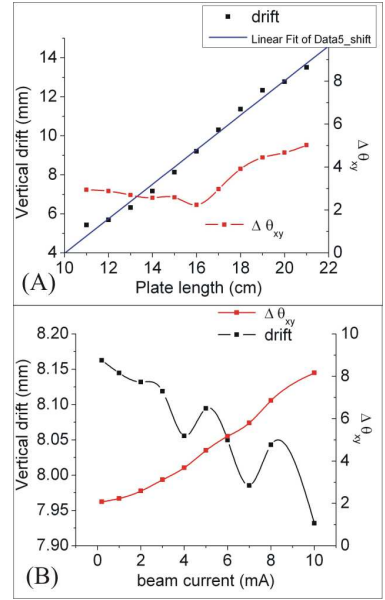

Figure 8: Effect of plate length (A) and beam current (B) on vertical drift and output angle of beam at 10keV. Plate length was 15*cm* for beam current study.

The output angle dependence of beam on the length of electric plates and beam current is shown in fig 8. The system was optimized for the beam energy of 10*keV*.

## CONCLUSIONS & OUTLOOK

We have presented the beam measurements and simulations along a single magnetic toroid. The investigations form the basis to study the beam dynamics in the whole ring. Further experiments are planned for investigations of coupled segments, the effect of trapped electrons and the beam injection scheme.